\documentclass[12pt,preprint]{aastex}
\newcommand{\emp}{\rm}

\newcommand{\be}{\begin{equation}}
\newcommand{\ee}{\end{equation}}
\newcommand{\bea}{\begin{eqnarray}}
\newcommand{\eea}{\end{eqnarray}}

\shorttitle{ Dark Matter Halos }
\shortauthors{Henriksen}

\begin{document}

\title{Nature and Nurture in  Dark Matter Halos}

\author{R. N. Henriksen}
\affil{Department of Physics, Engineering Physics and Astronomy,  Queen's University, Kingston, Ontario,Canada, K7L 3N6}
\email{henriksn@astro.queensu.ca}

\begin{abstract}
 Cosmological simulations consistently predict specific properties of dark matter halos, but these have not yet led to a physical understanding that is generally accepted. This is especially true for the central regions of these structures. Recently two major themes have emerged. In one, the dark matter halo is primarily a result of the sequential accretion of primordial structure (ie `Nature'); while in the other, dynamical relaxation (ie `Nurture') dominates at least in the central regions. Some relaxation is however required in either mechanism. In this paper we accept the recently established scale-free  sub-structure of halos as an essential part of  both mechanisms. Consequently; a simple model for the central relaxation based on a self-similar cascade of tidal interactions, is contrasted with a model based on the accretion of adiabatically self-similar, primordial structure.  We conclude that a weak form of this relaxation is present in the simulations, but that is normally described as the radial orbit instability.          
\end{abstract}

\keywords{cosmology:theory---dark matter---halo formation}

%%%%%%%%%%%%%%%%%%%%%%%%%%%%%%%%%%%%%%%%%%%%%%%%%%%%%%%%%%%%%%%%%%%%%%%%%%%%%%%%%%%%%%%%%%%%%%%%%%%%%%%%%%%%%%%%%%%%%%%%%%%%%%%%%%%%%%%%%%%%%%%%%%%%%%%%%%%%%%%%

\section{INTRODUCTION}

In Henriksen (2006b, H06; 2007, H07) a theory of dark matter relaxation has been proposed that is based on a temporally convergent series solution for the Distribution Function (true phase-space density or DF). The relaxation is effected in a non-mechanistic fashion by maximizing the local Boltzmann function calculated from this DF. This procedure determines all of the parameters in the DF and allows the density distribution $\rho(r)$ and the pseudo phase-space density $\phi(r)\equiv \rho(r)/\sigma(r)^3$ to be calculated, as well as any other quantity such as specific angular momentum. 
Studying these results one finds that, starting from the position and corresponding slopes found in the simulations, either the density flattens rapidly or the pseudo density steepens rapidly within the next decade of smaller radius. There is at present no convincing evidence for either of these trends in the simulations, although unfortunately this region is at or near the current resolution limit. This paper attempts to understand the proposed relaxation more intuitively by suggesting a simple mechanistic model. Such a view may be briefly summarized as `nurture'.

In Salvador-Sol\'e et al. (2007, S07) a rather different explanation for the simulated structures is proposed. In effect one uses the Press-Schechter (1974) formalism as expressed by Lacy and Cole (1993) to calculate the instantaneous merger rate. Since most of the halo mass is added by many small objects, this is approximated by being smooth. This smoothness allows one to accrete the dark matter halo from post-recombination large scale structure in a sort of `layer cake' fashion, where each layer may be deduced from the  accretion flow at the epoch when it was added. The innermost layers correspond to the earliest times so that `inside-out' growth is established.  The predicted density profile compares well to NFW (Navarro, Frenk and White, 1997; Navarro et al. 2004) profiles for various masses over appropriate ranges and even better over the whole range of scales to an Einasto (Navarro et al., 2004) or S\'ersic profile (e.g. S07). The latter profile has a finite density at the centre but an infinite slope. Similar comparisons may be made for velocity dispersion and angular momentum (Gonzales-Casado et al., 2007). Such a view may be briefly summarized as `Nature', although there is also a kind of relaxation that leads to adiabatic self-similarity (e.g. Henriksen 2006b, H07) in the asumption of ``smoothness''.

Unless it is also hidden in the asumption of `smoothness', the latter picture leaves little room for `thermodynamic relaxation' (ie maximum entropy) as an element in the simulations. Nevertheless, whether or not such relaxation is present in reality in order to explain possible  density cores remains an open question. Moreover the remarkable results of the `Via Lactea' simulation (Diemand, Kuhlen and Madau, 2006-DKM06; Madau, Diemand, and Kuhlen, 2008-MDK08) show elaborate sub-structure that is characterized by a definite mass spectrum. Such a hierarchy should be interacting tidally and by dynamical friction and so it provides a mechanism for relaxation (see e.g. the discussions in El-Zant et al., 2004, H07, H06 and  Henriksen, 2006a). 

One indication of the possible presence of such relaxation has been provided by Hoffman et al. (2007). They observe that the halo surface density at the NFW scale radius $r_s$  (that is $\rho_sr_s$) remains constant during the evolution of an individual halo. This together with virialization within $r_s$ allows them to conclude that $\phi_s\propto r_s^{-5/2}$. Some time ago, virialization together with strong dynamical inter-scale coupling were shown to be equivalent to constant surface density and virialization throughout a cascade of structures. This was in a completely different context (Henriksen and Turner, 1984; Henriksen, 1991 and references therein) where it was studied in the context of star formation. It was proposed that a hierarchy of molecular clouds slowly evolved by collisional and tidal interactions into the stellar Initial Mass Function (IMF) as the result of a kind of `ballistic turbulence'. The observation of Hoffman et al (2007) together with the power law substructure (DKM06) suggests a similar argument may apply to dark matter halos.

In the stellar case it has been difficult to find any evidence for such an initial fragmented state, and so ironically  the picture may be more relevant to dark halos than it is to luminous stars!  However Hoffman et al. have not applied the idea of a dynamical cascade to all scales below $r_s$. This will be the subject of the next section.
In the third section we give a naive calculation that imitates the layer cake approach of (S07), by simply assuming adiabatic self-similarity as dictated by the primordial perturbation spectrum. Finally in the conclusions we discuss the liklihood of either sort of relaxation. We conclude  that a weak form of cascade relaxation is present in the simulations that is not distinguishable from the radial orbit instability. We conclude further that the pure cascade evolution should lead to a sharp break in the  density and pseudo-density trends as simulated  currently in the next decade or so of resolution. Otherwise we are left only with the weak form of relaxation that is described in this paper as either `nature' or adiabatic self-similarity, and which seems to be equivalent to the radial orbit instability.
 
\section{Cascade Relaxation}

In this section we apply the self-similar model of cascade structure  introduced for molecular clouds in Henriksen and Turner (1984; HT84 subsequently) and summarized in Henriksen (1991; H91 subsequently). We describe first the `initial'  self-similar structure of the cascade and subsequently estimate its equilibrium radial structure using a local maximization of entropy (following Henriksen (2007)). 

\subsection{Initial Cascade}

We use a fractal description of the number of halo substructures according to 
\be
n(\ell,r_s)=(\frac{\ell}{r_s})^{-D}\equiv \frac{dN(\ell,r_s)}{d\ln{(\ell/r_s)}}.\label{fractal}
\ee

Here $n(\ell,r_s)$ is the number of substructures inside the NFW scale radius $r_s$ each having scale $\ell$, and $D$ is a cascade parameter that is usually described as the fractal index since it need not be integer. This is the true fractal view wherein the structures are separated by discrete steps in scale which, in a self-similar fractal, are effected by a constant scale ratio. In fact equation (\ref{fractal}) follows by recursion once the number of substructures of size $\ell$ in an object of size $L$ is assumed to be $n(\ell,L)=(\ell/L)^{-D}$.  The transition to the continuum description is also indicated in equation (\ref{fractal}). We set $n(\ell,r_s)$ equal to 
the derivative with respect to the logarithm of scale (so equal differences are equal ratios) of the cumulative number of substructures inside $r_s$ having scale less than $\ell$, namely $N(\ell,r_s)$. 

The description of the cascade used in this paper can appear confusing because of the intrinsic number of parameters and of several ad hoc notations that are used either for brevity or for correspondence with previous work. For the convenience of the reader we  assemble here the various notations in  table 1, even before their introduction in the text. There are really only two cascade parameters under our assumptions namely $\alpha$ and $\beta$ in terms of which all other cascade quantities may be expressed. These are defined below just before equation (\ref{MLMell}) and by equation (\ref{beta}). We use generally a systematic ad hoc notation wherein $D_x$ stands for the negative power of the dependence on radius of the quantity $x$. The symbol  $D$ without a subscript corresponds to the fractal index of the cascade elements as above.

\begin{table}[h]
\begin{center}
\caption{Definitions\tablenotemark{a}\label{table1}}
\begin{tabular}{lc}
\tableline\tableline
Parameter & Definition \\
\tableline
$D$ & $\,\,~$ $2+\beta$   \\
$D_\rho$ & $\,$ $1-(\alpha+\beta)$  \\
$D_\phi$ & $\frac{(5+\alpha+\beta)}{2}$ \\
$D_{j^2}$ & $-(3+\alpha+\beta)$  \\
$\nu$  & $3+2(\alpha+\beta)$   \\
$i_1$ & $\alpha$\\
$K$ & $\ln{(r_s/\ell_o)}$\\
$\ell_o$ & Cascade top scale\\
\tableline
\end{tabular}
%% Any table notes must follow the \end{tabular} command.
\tablenotetext{a}{Ad hoc quantities in terms of Cascade parameters}
%\tablenotetext{b}{Heliocentric velocity.}
\end{center}
\end{table}

We define the scale of each substructure as  a kind of gravitational correlation length (e.g HT84) over which the velocity dispersion $\sigma(\ell)$  and the total mass $M(\ell)$ satisfy the virial condition 
\be
\sigma^2(\ell)=\frac{GM(\ell)}{\ell}.\label{virial}
\ee
We show below (equation (\ref{F}) and subsequent remark) that a self-similar virial cascade of these sub-structures implies a hierarchical vector velocity that adds randomly to give a virialized  structure on the scale $r_s$, as it should. This local virialization assumption has been found to hold recently in a general class of distribution functions (Wyn Evans and An, 2005, An and Wyn Evans, 2005). This is reasuring since the assumption has been used intuitively for some time in these cascade models, as noted in the introduction to this paper. 

In order to parameterize  each substructure as  composed of substructures plus a smooth distribution of particles on the scale $L$, we introduce the fraction $f(\ell,L)$ as the fraction of mass $M(L)$ contained in structures of the next sub-scale $\ell$. In a self-similar fractal it is natural to take this as $f(\ell,L)=(\ell/L)^\alpha$, where  $\alpha$ is thus another parameter of the cascade. Consequently the mass of an object of scale $L$ is given by 
\be
M(L)=\frac{n(\ell,L)M(\ell)}{f(\ell,L)}.\label{MLMell}
\ee
 By recursion this yields the mass of an object of scale $\ell$, in terms of the total mass inside $r_s$, namely $M_s$,  as 
\be
M(\ell)=M_s(\frac{\ell}{r_s})^{(D+\alpha)}.\label{Mell}
\ee

the mean density of an object of size $\ell$ is therefore 
\be
\overline\rho(\ell)\equiv \frac{M(\ell)}{4\pi\ell^3/3}=\rho_s (\frac{\ell}{r_s})^{-(3-(D+\alpha))},\label{density}
\ee

where $\rho_s\equiv M_s/(4\pi r_s^3/3)$.

In H91, the mean density is taken to vary as $\ell^{-D_\rho}$ so that in the present notation 
\be
D_\rho=3-(D+\alpha),\label{Drho}
\ee
which may be employed where convenient. Moreover $\alpha$ is thereby seen to be the `aggregation index' $i_1$ as defined in H91. When this quantity equals zero, all of the mass at each scale is composed of identifiable substructures. We will refer to this condition as an `aggregation cascade' for brevity. In H07 the index $D_\rho\equiv 2a$.

We will, for the purposes of describing the simulations with this model, frequently want to make the correspondence between scale $\ell$ and spherical radius $r$. In the spirit of the dynamical coupling between scales advocated in HT84 and in H91 we suppose that this correspondence follows by equating the dynamical time at $r$ to the dynamical time of the maximum sub-structure of scale $\ell_r$ permitted at $r$, so that
\be
\frac{\sigma^2(\ell_r)}{\ell_r^2}=\frac{GM(r)}{r^3}.\label{coupling}
\ee
  Hence recalling equation (\ref{virial}) for the scale $\ell_r$, and using equation (\ref{Mell}) for a sphere of radius $r$ (rather than $r_s$) to obtain $M(r)=M(\ell_r)(\ell_r/r)^{-(D+\alpha)}$, we conclude that 
\be
(r/\ell_r)^{(D+\alpha-3)}=O(1), 
\ee
and hence $\ell_r\approx r$. This means that the maximum scale at any $r$ is about equal to $r$ itself. 

% Interestingly, the parallel/perpendicular mode coupling condition of equation% (\ref{coupling}) is similar to that used in the Goldreich-Sridhar (1995) theo%ry of MHD turbulence as remarked in Cho, Lazarian and Vishniac (2002).

The first constraint on such a hierarchy from the simulations comes from the DMK06 and MDK08 papers where they present evidence for the substructure cumulative mass function $N(M_{sub})$ in the form 
\be
\frac{dN(M_{sub})}{d\ln{M_{sub}}}\propto M_{sub}^{-q},\label{obsmass}
\ee
where $q$ is approximately in the range $0.9<q<1.0$. In our model $M_{sub}\equiv M(\ell)$ and by equation (\ref{Mell}) $d\ln{M(\ell)/M_s}=(D+\alpha)d\ln{\ell/r_s}$ so that by substituting into the continuum form of equation (\ref{fractal}) we have our prediction for this result
\be
\frac{dN(M(\ell))}{d\ln{(M(\ell)/M_s)}}=\left(\frac{1}{D+\alpha}\right)\left(\frac{M(\ell)}{M_s}\right)^{-\frac{D}{D+\alpha}}.\label{mfunct}
\ee

It is therefore already clear that such a model can only fit the simulations if $\alpha<<1$. This would imply an approximate aggregation cascade (i.e. $i_1=\alpha\approx 0$) at least at the larger scales. This is not consistent with the $z=0$ simulation state (DMK06, MDK08), wherein they find that the sub-structure is only about five  percent of the total mass inside the virial radius.  However we must remember that at present we are describing the early state of the cascade where such a condition may have been  more appropriate. More appropriate because although the mass function does not seem very sensitive to epoch in the simulations the mass fraction in clumped material increases with $z$ (MDK08). 

%Less constraining because, as we shall see below, from the `Nature' perspectiv%e $D=n+3$ where  $n$ is the current scale power spectrum index ($P(k)\propto k%^n$) that increases to higher $z$. The larger $D$ is, the larger is the permit%ted $\alpha$. On very large current scales (and hence early times)  $n=1$ so t%hat $D=4$.

The key to the subsequent evolution of such a cascade is the interscale dynamical coupling, which on our view promotes the relaxation of the system (HT84,H91)  and is due implicitly to tidal interactions, which we take to include collisions with all impact parameters. This is quantified by comparing the dynamical time on a scale $L$, namely $t_{dyn}=L/\sigma(L)$ to the collision time of the next internal subscale $\ell$, namely $t_{coll}=1/((n(\ell,L)/L^3)\pi\ell^2\sigma(L))$. A strongly coupled cascade would have $t_{coll}/t_{dyn}\approx 1$, but we generalize slightly by introducing a third parameter $\beta$ such that
($\pi$ is introduced for convenience)
\be
\frac{t_{coll}}{t_{dyn}}=\pi(\frac{\ell}{L})^\beta,\label{beta}
\ee
so that by using the various definitions

\be
\frac{L}{\sigma(L)}(\frac{\ell}{L})^\beta=\frac{L}{(\frac{\ell}{L})^{(2-D)}\sigma(L)},
\ee
or finally
\be
(\frac{\ell}{L})^{(2-D+\beta)}=1.
\ee
 
Since this can only hold for a scale ratio unequal to $1$ if the power on the left vanishes, our cascade  parameters are again reduced to two (i.e. $\alpha$ and $\beta$) as now we require 
\be
D=2+\beta,\label{D}
\ee
and hence 
\be
D_\rho=1-(\alpha+\beta).\label{DDrho}
\ee
If $\beta$ is negative the cascade becomes increasingly collision dominated at the small scales (see equation (\ref{beta})) and we expect the relaxation to be stronger there. Ultimately the sub-structure may be expected to merge into a continuum at small scales.
 
The pseudo phase-space density (Taylor and Navarro, 2001) may also be calculated in the cascade as $\overline\rho(\ell)/\sigma(\ell)^3$ from equations (\ref{density}, \ref{virial}, \ref{Mell}) to be 
\be
\phi=\phi_s(\frac{\ell}{r_s})^{-(5+\alpha+\beta)/2}.\label{phi}
\ee
Here $\phi_s\equiv \rho_s/\sigma_s^3$ and recall that we may expect to replace $\ell$ with $r$ on average. For brevity we may refer to $(5+\alpha+\beta)/2$ as simply $D_\phi$.

In the  strongly coupled, aggregation cascade ($\beta=0$, $\alpha=0$), this varies like $r^{-2.5}$ while $\rho\propto r^{-1}$ (ie constant surface density, HT84). These relations should hold everywhere in the un-evolved cascade, not just at any particular radius such as the NFW scale radius (Hoffman et al 2007). 

However this simple description does not fit the $z=0$  state of the simulations. This state requires that at a radius somewhat smaller than the NFW scale radius we have approximately $\alpha+\beta\approx -1$, so that $D_\rho\approx 2$ and $D_\phi\approx 2$. In addition $\alpha\approx 0$ in order to fit the sub-structure mass function . As the simulations are continued to smaller radii (down by about two orders of magnitude) $D_\rho$ decreases to about $1$ and slowly thereafter (Navarro et al., 2004), $D_\phi$ appears to remain close to $2$ (Austin et al. 2005; Dehnen and McLaughlin, 2005) and $\alpha$ remains close to zero to the resolution limit (DKM06).

 We do not expect that relaxation by cascade is relevant much beyond the NFW scale radius where the density profile is steeper that $r^{-2}$ and the `nature' of the environment probably predominates (see e.g. below and Henriksen and Widrow, 1999). In contrast, the evolution toward smaller scales should indeed manifest relaxation by this clump-clump mechanism. We thus turn next to examining whether or not this simple cascade model can evolve to smaller radii as expected from the simulations.
 
Such an examination was carried out also in  H07, but the calculations were rather formal. They were based on a renormalized, `adiabatically self-similar', distribution function together with the maximization of the local entropy density. It is this assumed maximization that takes the place of the detailed dynamics of the tidal interactions. In the next section we give a simpler version of this discussion based on the preceding cascade structure. Similar conclusions to those of H07 are reached nevertheless.

\subsection{Cascade Evolution}
 
The actual dynamical implementation of a  cascade of tidal interactions or collisions escapes a precise description for the moment. To avoid this impasse, we use the idea that these processes move the cascade through a succession of local equilibria (e.g. H07). Each equilibrium on a scale $\ell$ is taken to be associated with a maximum of the Boltzman H function. In Henriksen (2007) this was discussed in terms of parameterized distribution functions, whose parameters varied with scale. We attempt here to give a simpler version of the model in terms of the sub-structure cascade detected in the simulations.

As a result of this assumption the details of the interactions are ignored. They enter implicitly  however through equation (\ref{beta}). Should $\beta >>1$, then tidal collisions and dynamical friction would not play a role in the dynamical evolution below $r_s$. However our evolution below gives the range of $\alpha+\beta$ and together with $\alpha$ small this range requires $|\beta|$ to never be far from unity in order to approximate the simulations. And in fact we shall see that the relaxation detected in the simulations requires $\beta <0$. 
    
In Henriksen (H91) it was suggested that the cascade evolved on a given scale when a dimensionless time $\tau(t,\ell)$ reached a critical value $\tau_J$ on that scale. In the stellar case this heralded the transformation from molecular cloud sub-structure to a pre-stellar `core'. However this depended essentially on dissipation, which is not present in the present application. In this context we expect the parameters of the initial cascade to be modified on each scale at some dimensionless time due to tidal transfer of energy and angular momentum. We expect this process to go faster on the smaller scales and to transfer energy, angular momentum and mass to larger scales.

As a simple implementation of this last remark, in order to provide some expectation of a timescale for the equilibrium calculation to follow, we may estimate the form of $\tau_J$  by dimensional analysis of the initial cascade. Given a cascade characterized only by power laws in radius for both the density and the pseudo-density, their  dimensional factors (say $\lambda$, $\mu$ respectively) may be taken as the dimensional constants that define the cascade self-similarity. These may be combined with the gravitational constant to yield two expressions for a local time scale, namely from the density as $\tau_\rho=\sqrt{
G\rho(\ell)}t\equiv \sqrt{G\rho_s}(\ell/r_s)^{-D_\rho/2}t$ and from the pseudo-density as $\tau_\phi=(1/(G\phi(\ell)\ell^3)t\equiv 1/(G\phi_sr_s^3)(\ell/r_s)^{-(3-D_\phi)}t$. The dimensional constants have been written here implicitly in terms of the values of density and pseudo-density at $r_s$, namely $\lambda =\rho_s r_s^{D_\rho}=\rho(\ell)\ell^{D_\rho}$ and $\mu=\phi_s r_s^{D_\phi}=\phi(\ell)\ell^{D_\phi}$. In order to obtain a single estimate of the evolutionary time scale we would like these two estimates to agree. Happily this requires $D_\phi=3-D_\rho/2$, which is in fact identically satisfied by their definitions in terms of cascade parameters $\alpha$ and $\beta$. They also agree numerically under our conditions. 

Other possible time scales for the cascade, such as the global dimensionless dynamical time $(\sigma_s/r_s)t$, are not independent of these dimensional factors. This latter dynamical time is in fact just $\sqrt{G\rho_s}t$, which is $\tau_\rho$ above at the scale $r_s$.

Setting  the common $\tau$ defined above equal to a  value $\tau_J$ independent of scale estimates  the initial evolution of the  cascade. We see that the scale of the evolved cascade increases with a positive power of $t$, the age of the system. Hence, descending in scale, we expect to find an increasingly evolved cascade at a fixed $t$. The numerical value of $\tau_J$ may be estimated to be of order unity since it is numerically equal to $\sqrt{G\rho_s}t$ and we expect the relaxation to occur in several dynamical times on the scale $r_s$. Thus  the time (from the origin of the cascade) to significant evolution on a scale $\ell$, say $t_J$, is given by
\be
t_J(\ell)=\frac{\tau_J}{\sqrt{G\rho_s}}(\frac{\ell}{r_s})^{D_\rho/2}.\label{tJ}
\ee
Hence as either we descend the sub-structure scale, or as $D_\rho$ decreases in time (see below), the relaxed state becomes less dependent on scale. Such a picture agrees with  equation (\ref{beta}) when the initial $\beta$ is negative.

The preceding estimate can only be approximate as it does not  take into account the adiabatic evolution of the cascade parameters. To achieve this, as remarked above, the actual dynamics of this evolution is replaced  by a statistical argument based on maximizing the local entropy density (Boltzmann H function: see e.g. H07 and discussion above). By applying  it locally in radius we hope to find the radial dependence of the evolution in the cascade parameters $\alpha$, $\beta$. In this connection we should recall the work of Hansen et al. (2006) who studied the velocity distribution in dark matter simulations, as well as similar work by Merrall and Henriksen (2003). In these papers evidence for relaxation (near Gaussianity and universality) is found in colliding and collapsing systems. In the Merrall and Henriksen paper, the radial velocity distribution function is fitted to a Gaussian (with distortions) under many conditions, while in Hansen et al both the radial and tangential velocities are fitted to a Tsallis entropy distribution. Some anisotropy is detected in the latter paper that argues against complete thermodynamic relaxation, but relaxation is still present. This anisotropy is liable to be dependent on net angular momentum in the system which tends to slow the approach to equilibrium. We ignore this effect in the nurture calculation, but not in the nature calculation, where the relaxation is indeed slower. 

To proceed we estimate the phase-space volume $\Omega$ available to substructures of scale $\ell$ inside the sphere of radius $r$ as (for distinguishable particles)
\be
\Omega(\ell,r)\approx (\frac{h^3}{\sigma_{cm}(\ell)^3r^3})^{n(\ell,r)}.
\ee
 Here $n(\ell,r)$ is the number of sub-structures on scale $\ell$ inside a sphere of radius $r$, $\sigma_{cm}(\ell)$ is the average velocity dispersion of the corresponding element relative to the centre of mass of the halo and $h^3$ is a convenient element of phase space with which to scale the phase space volume.

It is in fact appropriate to take $h^3$ as a volume close to the largest scale in the cascade, $r_s$, since the smallest scale is not well determined. This leads to the inverted form for the phase-space volume above in order to keep the entropy positive. Thus we write $h=\ell_o \sigma_o$ where $\ell_o$ and $\sigma_o$ are quantities on a scale close to the top of the cascade. Using equations (\ref{virial}, \ref{Mell}) and the various definitions 
\be
\sigma(\ell)=\sigma_s (\frac{\ell}{r_s})^{\frac{1+\alpha+\beta}{2}},\label{ssigma}
\ee
so that we obtain $h$ by applying this last result to $\ell_o$ and then multiplying by $\ell_o$ as  
\be
h=r_s\sigma_s(\frac{\ell_o}{r_s})^{\frac{3+\alpha+\beta}{2}}.
\ee
We expect $\ell_o$ to be somewhat smaller than $r_s$ to form the top of the cascade. The advantage to this choice for $h$ over simply the product of $r_s$ and $\sigma_s$ is that it allows for the evolution of the cascade with  varying parameters $\alpha$ and $\beta$. We find below that  $h$ increases during the evolution so that the cascade indeed grow from small scales to large scales.
%\end{document}

We calculate $\sigma_{cm}(\ell)$ as the result of  $N(\ell)$ steps (not to be confused with the cumulative mass function) of a vector random walk between scales $\ell$ and $r_s$, each step of  size ${\cal R}$. We should first note that equation (\ref{Mell}) can be written in terms of the $N^{th}$ step   as 
\be
M(N)=M_s{\cal R}^{N(D+\alpha)},\label{MN}
\ee
where $N$ is related to $\ell$ by
\be
N(\ell)=\frac{\ln{\ell/r_s}}{\ln{\cal R}}.
\ee
Then we write the halo centre of mass velocity on the scale $N$ namely $\vec{v}_N$ as
\be
\vec{v}_N=\vec{v}_{N-1}+\Delta\vec{v}_N,
\ee
where $\Delta\vec{v}_N$ is the randomly oriented velocity step  from $N-1$ to $N$.  We take this once again to have amplitude equal to the virial velocity on that scale, namely $\sqrt{GM(N)/{\cal R}r_s}$. Hence by squaring and averaging over all directions so that we may drop the cross term we obtain 
\be
\sigma_N^2=\sigma_{N-1}^2+(\Delta\vec{v}_N)^2,
\ee
 and so by recursion to $r_s$ as the largest scale where $N=0$
\be
\sigma_{cm}^2(N)=\sigma_s^2\left(1+\Sigma_{j=1}^{j=N}~{\cal R}^{j(D+\alpha-1)}\right).
\ee
 We have set $GM_s/r_s=\sigma_s^2$.

On summing the preceding geometric series we find finally  

\be
\sigma_{cm}(\ell)^2=\sigma_s^2 F^2({\cal R},\alpha,\beta),\label{vcm}
\ee
where we use the ad hoc definition 
\be
F^2({\cal R},\alpha,\beta)\equiv\left(1+{\cal R}^{(1+\alpha+\beta)}(\frac{1-{\cal R}^{(1+\alpha+\beta)N(\ell)}}{1-{\cal R}})\right).\label{F}
\ee
Note that on setting $N(r_s)=0$ we obtain $F=1$ and thus equation (\ref{vcm}) merely restates the virialization on the scale $r_s$.  

We will normally choose ${\cal R}=1/3$ arbitrarily in our examples, since changing this value merely changes the scale associated with a given $N$.

The mean population $n(\ell,r)$ is estimated as $n(\ell)P(\ell,r)$ where $P(\ell,r)$ is the probability that a structure of scale $\ell$ should be found inside a sphere of radius $r$. We write this as 
\be
P(\ell,r)=\frac{\exp{-(\frac{M(\ell)\sigma_{cm}(\ell)^2}{2GM(r)^2/r})}}{Z},
\ee
%\end{document}
 where the `partition function' $Z$ is given approximately by the integral
\bea
Z({\cal R},\alpha,\beta)=& \frac{1}{r_s}\int_0^\infty~dre^{-\frac{rM(\ell)\sigma_{cm}(\ell)^2}{2GM(r)^2}}\nonumber\\ 
&\equiv \frac{1}{\nu}\int_1^\infty~du~u^{-\frac{\nu+1}{\nu}}e^{-C^2u}.
\label{partition}
 \eea
 We use the ad hoc notation 
\be
C^2\equiv\frac{ F^2(\nu, {\cal R})}{2}{\cal R}^{\frac{N(\nu+1)}{2}}.\label{C}
\ee
and $\nu$ is a convenient combined cascade parameter defined as 
\be
\nu=3+2(\alpha+\beta).
\ee
The convenient variable $u \equiv (r/r_s)^{-\nu}$.

There is a degeneracy between the parameters $\alpha$ and $\beta$ as they only appear below as a sum or equivalently as $(\nu-3)/2$.
The partition function can be evaluated in terms of Whittaker functions, but we shall not need this in the first approximation.

We now construct a local `entropy' for the gas of sub-structures $\ell$ in the sphere of radius $r$ as 
\be
S(\ell,r)=k\ln{\Omega(\ell,r)},\label{entropy}
\ee
and by assuming that this attains a maximum as a function of the cascade parameter $\nu$ and $x\equiv r/r_s$ for a fixed scale (equivalently $N$) we obtain the equation (cf H07)
\be
\frac{d\nu}{dx}=-\frac{\partial_x\ln{S}}{\partial_\nu\ln{S}},
\ee
which computes the evolution $\nu(x;N)$.
Explicitly this  takes the form
\be
\frac{d\nu}{d\ln{w}}=\frac{3e^{-w\nu}+C^2\nu A^*}{(3/2)e^{-w\nu}(\partial_\nu\ln{F^2}+K/2)-(C^2w+\partial_\nu C^2+e^{-w\nu}(\frac{N}{2}\ln{\cal R}+\partial_\nu\ln{Z}))A^*}.\label{evolution}
\ee
In this equation we use the previous `ad hoc' notation plus another as 
\be
A^*(x;R,\nu)\equiv -3w+\frac{3}{2}\ln{F^2}+\frac{3}{4}K(\nu+3),
\ee

where $w=-\ln{x}$, and $K=\ln{(r_s/\ell_o)}$.

Equation (\ref{evolution}) is a rather complicated differential equation for $\nu(x;N)$ in which a singular point ($0/0$) may appear. 
Fortunately, except possibly at the singular point, we can simplify it considerably by realizing that the derivative of the logarithm of $Z$ is small and that $F^2$ is never very far from $1$ for $\nu\ge 1$. Adopting $F^2=1$ also simplifies $C^2$ substantially. This gives us a working equation in the form (\ref{evolution}) but without the terms in 
$\ln{Z}$ and $\ln{F^2}$. Moreover from equations (\ref{C}, \ref{F}) we have  
\be
C^2\approx\frac{1}{2}\exp{(\frac{\nu+1}{2}N\ln{\cal R})}.
\ee

An examination of the singular point in the simplified equations shows that in general it is a focal point and solutions can not be expected to pass smoothly through it. It seems unlikely that adding the small terms omitted above will change this behaviour.

It transpires that the initial condition that we wish to apply to  equation (\ref{evolution}) is $\nu=1$ at $w=0$. This ensures that near $r_s$ we have $\alpha+\beta=-1$ so that $D_\rho=D_\phi=2$ as simulated. Since we expect $\alpha$ to be small, we can assume that $\beta\approx -1$. This means that the initial cascade is `uncoupled' at larger scales (cf equation \ref{beta}).

Equation (\ref{evolution}) allows us in principle to find how each scale ($N$ or $\ell$) evolves to smaller $r$, as the local entropy attains its maximum. Should $\nu$ increase towards $5$ or $\alpha+\beta=+1$, then a density core has formed ($D_\rho=0$) at that scale and the corresponding radius. If this evolution occurs before encountering the singular point, then our approximations are justified. Moreover because of the focal nature of the singular point, there is unlikely to be a unique critical curve that gives the `physical' behaviour.

\begin{figure}[ht!]
\begin{tabular}{cc} %This will make a two-column figure
\rotatebox{0}{\scalebox{.4} %change the angle and scale as you need
{\includegraphics{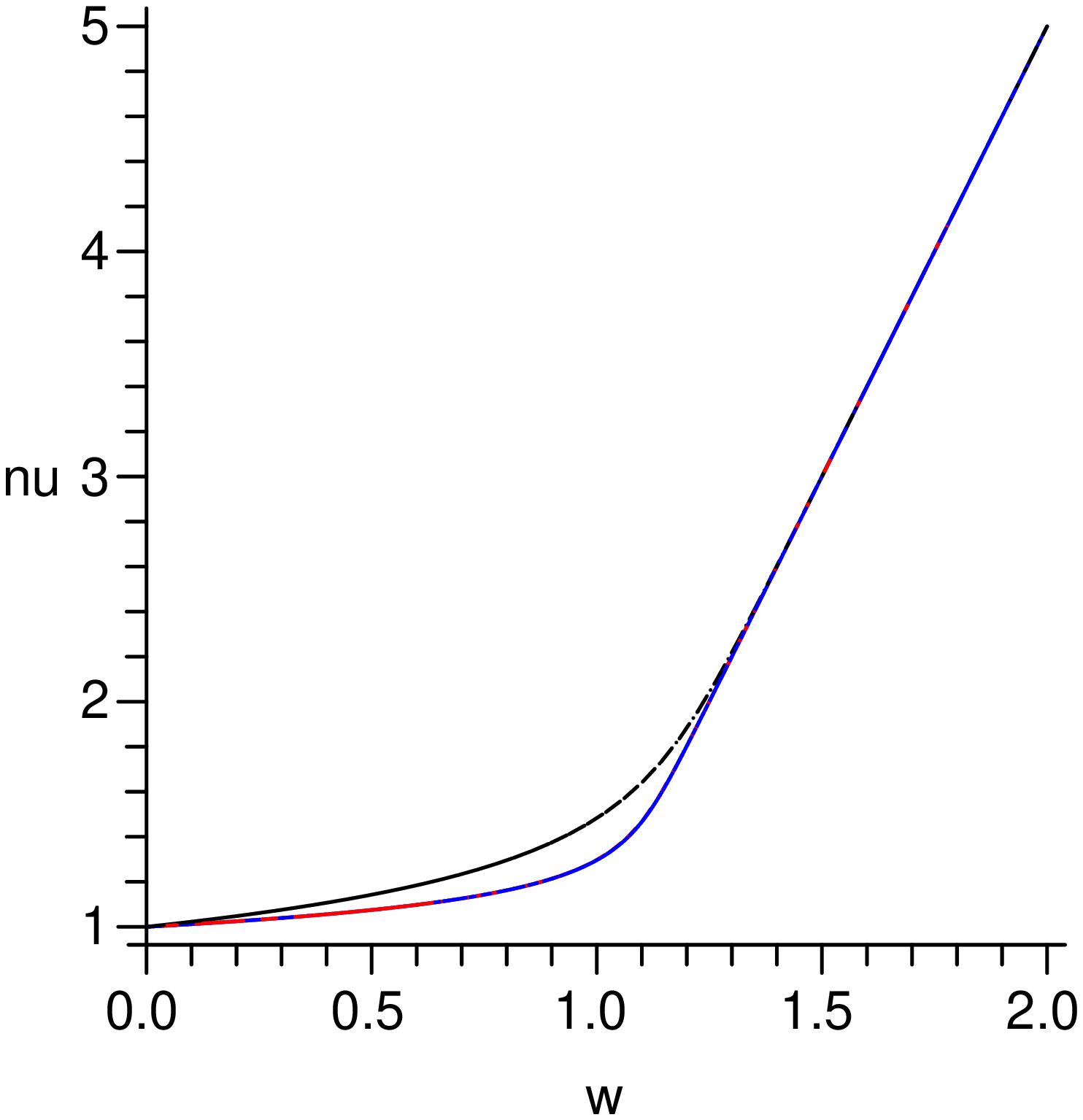}}}&
\rotatebox{0}{\scalebox{.4} %change the angle and scale as you need
{\includegraphics{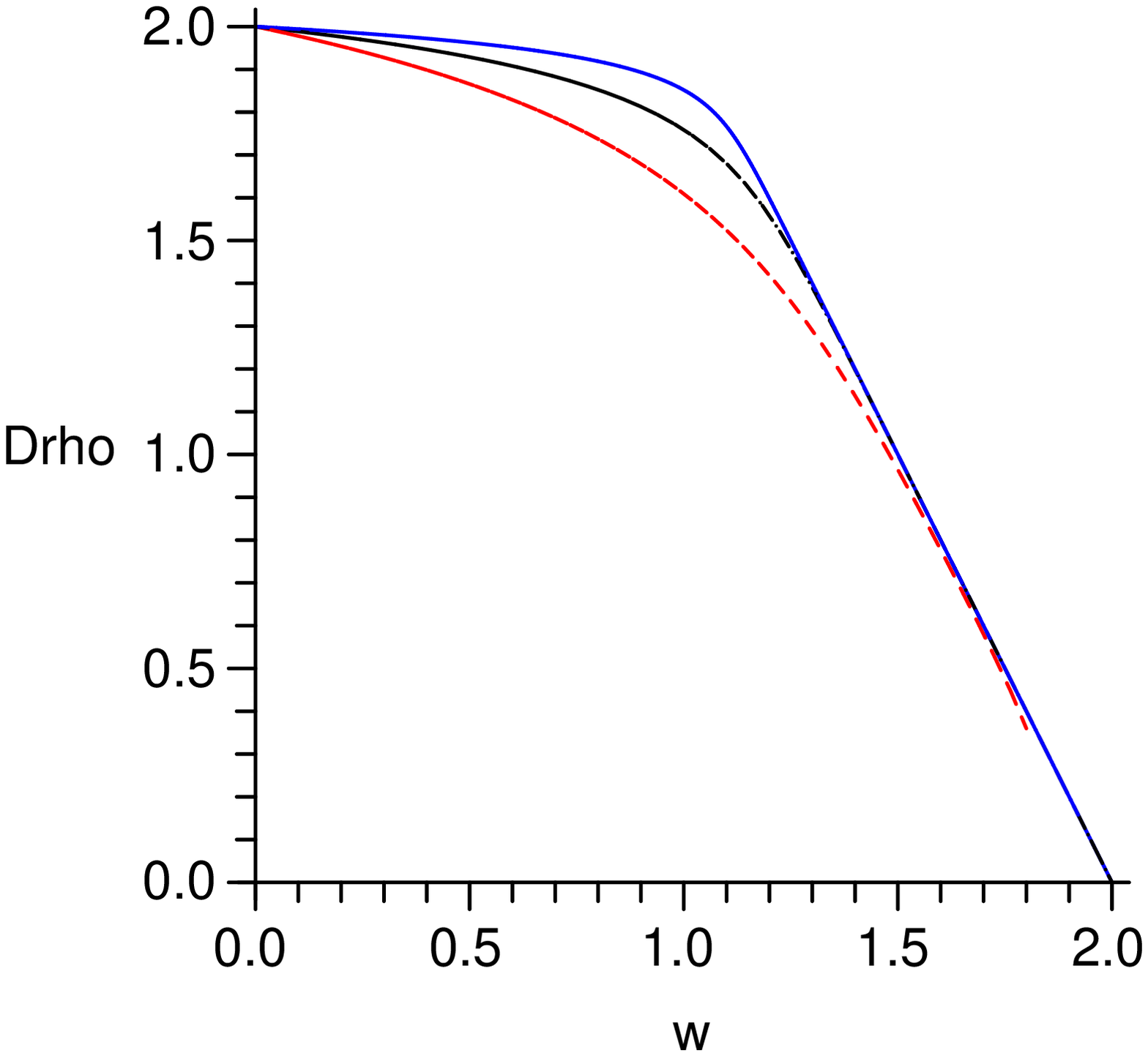}}}\\
\rotatebox{0}{\scalebox{0.4}
{\includegraphics{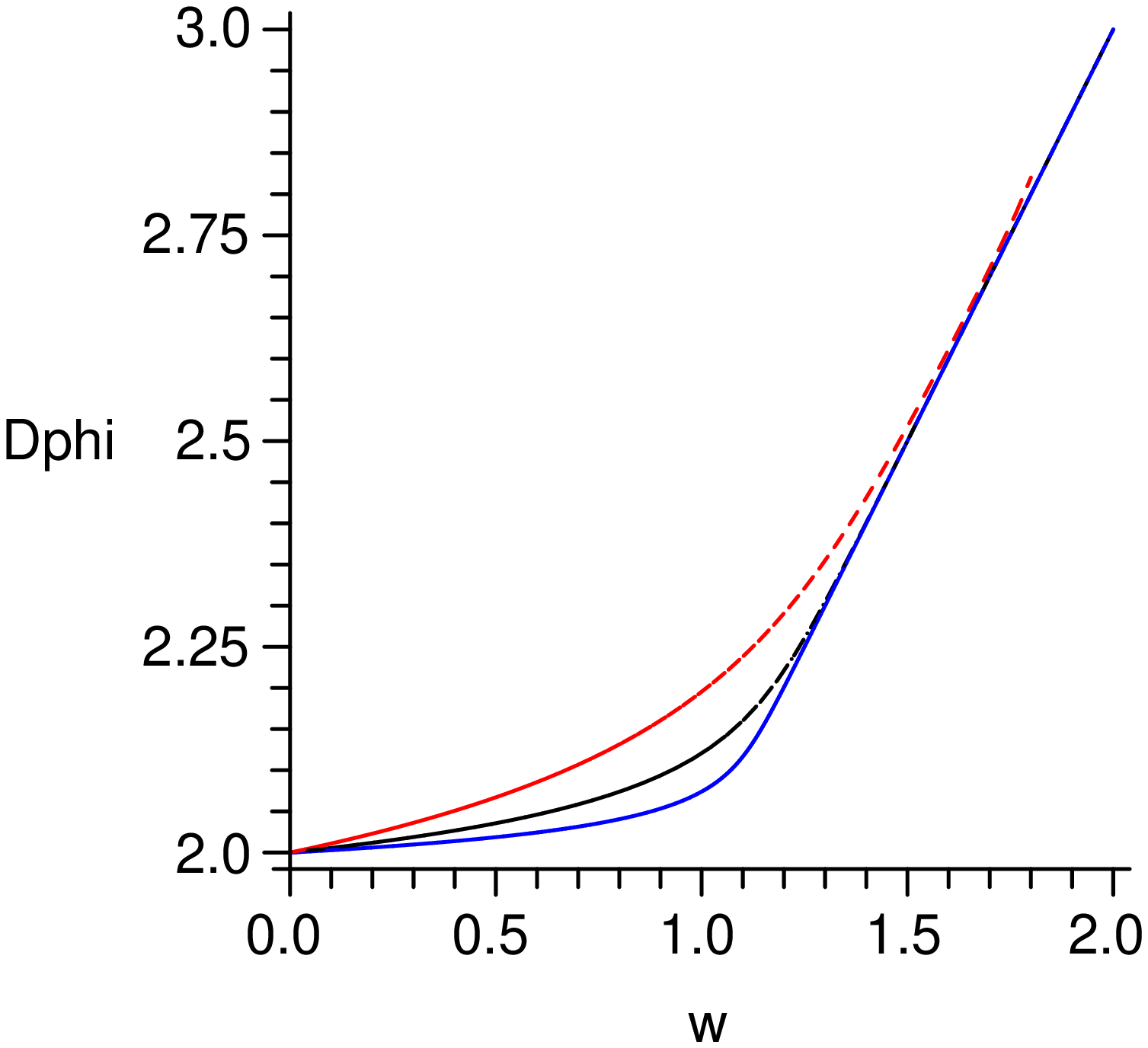}}}&
\rotatebox{0}{\scalebox{0.4}
{\includegraphics{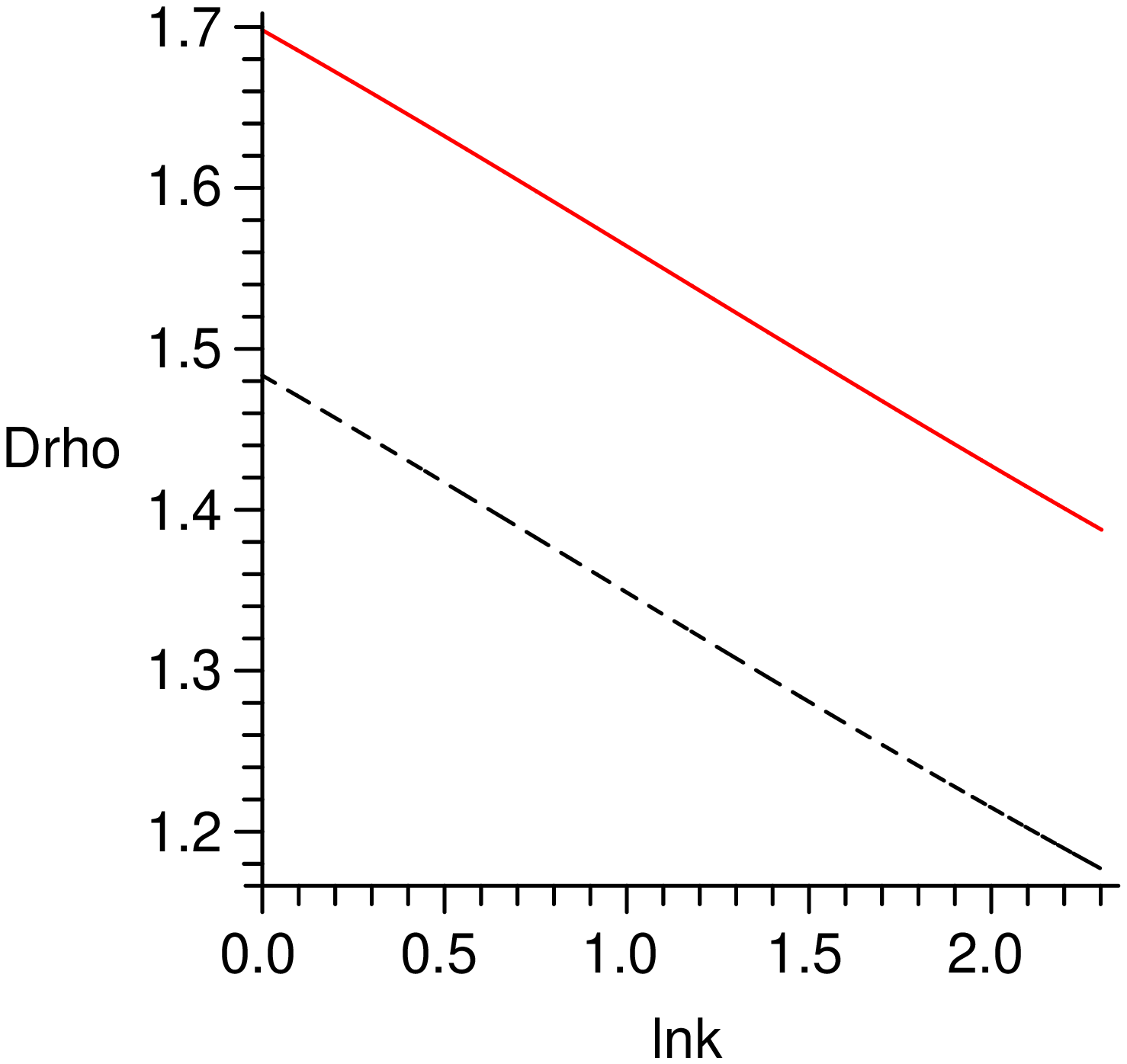}}}\\
\end{tabular}
\caption{The left panel shows the evolution of $\nu$ with log of the reciprocal radius for three different scales. The top line is for $N=4$, while the bottom line gives the behaviour both for $N=8$ and $N=2$. The right panel shows the corresponding density slope $D_\rho$ with $N=8,4,2$ respectively from the top curve to the bottom curve. The bottom left panel shows $D_\phi$ with $N=2,4,8$ respectively from the top curve to the bottom curve. All cases have ${\cal R}=1/3$ and $K=1$. The bottom right panel gives $D_\rho$ from the `nature' calculation below for $\delta=5$ (dashed line) and $\delta=4.5$ solid line. }
\label{fig:evolve}
\end{figure}

We see in figure (\ref{fig:evolve}) that $\nu$ does indeed evolve in this fashion over a range of scales.  As was  found in H07, this behaviour also implies that $D_\phi$ takes on the value $3$ as the core flattens.

More particularly in figure (\ref{fig:evolve}) we see that $\nu$ for all scales remains close to its  initial value while $w$ increases by unity, after which it rapidly converges to universal linear behaviour in $w\equiv -\ln{r/r_s}$. This behaviour requires the negative of the density slope $D_\rho$ to pass through the value $1.5$ in the same interval, after which it declines linearly with $w$ for all scales. The negative pseudo-density slope $D_\phi$ rises slowly over the same interval (more rapidly for the top of the cascade) after which it increases linearly with $w$ for all scales. This logarithmic behaviour of $D_\rho$ and $D_\phi$ was also found in H07 and was shown there to be a reasonable fit to the empirical results (Navarro et al, 2004)over a limited range in $r$. Ultimately at smaller $r$ it appears to vary too rapidly relative to the simulation. 
 
\begin{figure}[ht!]
\rotatebox{0}{\scalebox{0.5}
        {\includegraphics{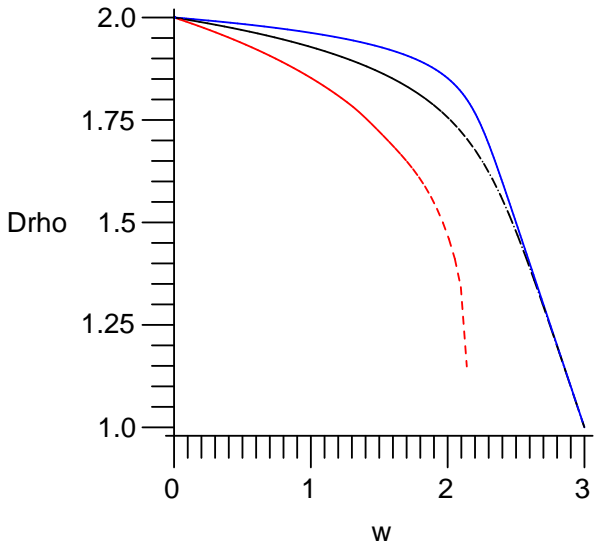}}}
\caption[]{\label{fig:K2evolve} 
This figure shows the same cases as found in the upper right panel of figure (\ref{fig:evolve}) but with the value $K=2$. One sees that for $N=2$ (bottom curve), which is now at the top of the cascade, the focal point behaviour is encountered. The curves for $N=4$ and $N=8$ (middle and top curves) are substantially slowed relative to the $K=1$ behaviour.}
\end{figure}

We note that this behaviour $\nu(x)$ is all on one side of the singular point of the differential equation, but such solutions yield the physical regime for our parameters.In fact only for the case $N=2$ (see e.g. figure (\ref{fig:K2evolve}) does the focal point fall into the interesting region $1<\nu<5$, when it occurs at $w\approx 1.8335$, $\nu\approx 2.2273$ (where the slope of the curve $D_\rho$ becomes infinite). The variation in $\nu$, and hence  in $D_\rho$ and $D_\phi$, can be slowed by reducing the upper scale of the cascade (increasing $K$), but then the singular `region'  may be reached  before $\nu =5$ as for $N=2$ above. This behaviour is indicated in figure (\ref{fig:K2evolve}). We see that as the top of the cascade is reduced in size , the smaller scales (larger $N$) slow down their evolution substantially. Indeed it would be difficult to distinguish them from the nature case for $n=4.5$ shown in the lower right panel of figure(\ref{fig:evolve}) over a limited range in scale. 

%\begin{figure}[ht!]
%\begin{tabular}{cc} %This will make a two-column figure
%\rotatebox{0}{\scalebox{.5} %change the angle and scale as you need
%{\includegraphics{DrhoK2.eps}}}&
%\rotatebox{0}{\scalebox{.4} %change the angle and scale as you need
%{\includegraphics{%}}}\\
%\rotatebox{0}{\scalebox{0.4}
%{\includegraphics{%}}}&
%\rotatebox{0}{\scalebox{0.4}
%{\includegraphics{%}}}\\
%\end{tabular}
%\caption{This figure shows the same cases as found in the upper right panel of figure (\ref{fig:evolve}) but with the value $K=2$. One sees that for $N=2$ (bottom curve), which is now at the top of the cascade, the focal point behaviour is encountered. The curves for $N=4$ and $N=8$ (middle and top curves) are substantially slowed relative to the $K=1$ behaviour. }
%\label{fig:K2evolve}
%\end{figure}

Although the preceding suggests that the interacting cascade starts well below $r_s$ in order to agree with the simulations,
the qualitative  agreement between this simple model for cascade relaxation and the predictions of the renormalized series for the distribution function ( H07) is marked. We have simply added here an explicit mechanism by which the local entropy may maximize, inspired by the structure cascade found in DKM06. 

Our conclusions in this section are moreover the same as in H07. If relaxation can occur in such a fashion as to maximize the local entropy, then there should be a flattening density core and a steepening pseudo-density in relaxed dark matter halos. Moreover this should happen relatively rapidly for scales near the top of the cascade, although this upper limit to the relaxed cascade may be on a substantially smaller scale than $r_s$  since it develops first at small scales. This allows the model to agree with the current simulations over a limited range.
We observe further that the likely evolutionary path indicated by figure (\ref{fig:K2evolve}) is to descend the largest scale curve ($N=2$) until the infinite
slope in the density index is reached ($w\approx 2$). At this stage the radial scale has been reduced so that the next  curve to the right in the figure becomes relevant by a horizontal displacement. This continues until the universal behaviour is encountered at small scales, which is rapid compared to the `nature' results also shown in figure (\ref{fig:evolve}).

In the next section we turn to an analytic approximation  to the  `nature' argument  based on adiabatic self-similarity (H06) for the relaxation and smooth accretion.

\section{Dark Matter Halos from Cosmological Accretion}

In Manrique et al. (2003), Zhao et al. (2003) and Salvador-Sol\'e et al. (2007; S07) the importance for the structure of dark matter halos of smooth accretion from cosmological conditions was emphasized. Indeed it was shown, particularly in the latter paper, that the assumption of smooth `inside-out' accretion based on the accretion of small objects from the Press-Schecter paradigm (Lacey and Cole, 1993) allowed the simulated structure to be reproduced (i.e. NFW or S\'ersic density profile, $M_s(r_s)$ correlation) over an impressive range of scales.

The key idea in these studies is that the density profile adjusts to  current  
accretion conditions in a sequential `layer-cake' fashion where the inner regions are fossil records of early cosmological times (`inside out' accretion). 
 Moreover, whenever there is a reasonable approximation to a power-law spectrum for the cosmological perturbations of the form $P(k)\propto k^n$, S07 found that the usual models for self-similar relaxation (Fillmore and Goldreich, 1984; Hoffman and Shaham, 1985; Henriksen and Widrow, 1999; Le Delliou and Henriksen , 2003) do convert the cosmological accretion (Lacey and Cole, ibid) into the simulated density according to (our notation, with $n$ an index independent of the previous section)
\be
D_\rho\equiv 2a=\frac{3(3+n)}{\delta+n}.\label{an}
\ee
Here $\delta=4$ around a local maximum (Hoffman and Shaham, 1985) and $\delta=5$ around an $n\sigma$ peak (Henriksen  H06). 

Thus even in this picture that relies directly on cosmological `nature', there is some relaxation occurring. This has been described as ``relaxation in an accretion bath'' by Salvador-Sol\'e (private communication) but in the language of Henriksen (H06) this may be considered as a kind of `adiabatic self-similarity'. That is, the global constants determining the self-similarity are dynamically changing. In the original picture (H06) this is due to local dark-matter relaxation, which we described in H07 and in the previous section by maximizing the local entropy. Here it is due rather to the changing cosmological infall with scale $k^{-1}$ (and hence cosmological time) that dictates the local self-similarity through equation (\ref{an}). Since the self-similarity is itself a result of relaxation, this is really a mixture of `nature' and `nurture' as remarked above.

Just as in the previous discussion of Cascade relaxation, we include a table that defines the central quantities of this section.
\bigskip

\begin{table}[h]
\begin{center}
\caption{Definitions\tablenotemark{a}\label{table2}}
\begin{tabular}{lc}
\tableline
Quantity & Definition \\
\tableline
$D_\rho$ & $\,\,$~ $3(3+n(k))/(\delta+n(k))$   \\
$P(k)$ & $\,$  Power Spectrum see equation(\ref{PS})  \\
$n(k)$ & adiabatic spectral index see equation(\ref{nindex}) \\
$k$ & structure wave number $\approx 1/r$ \\
$Q/Q^*$  & $(2+B/A^2)^2$   \\
$B,A$ & Power Spectrum numerical scales\\
$a(k)$ & $D_\rho/2$\\
%$\ell_o$ & Cascade top scale\\
\tableline
\end{tabular}
%% Any table notes must follow the \end{tabular} command.
\tablenotetext{a}{Adiabatic Nature quantities }
%\tablenotetext{b}{Heliocentric velocity.}
\end{center}
\end{table}

\bigskip

\newpage
Now in fact the usual approximation to a cold dark matter (CDM) power spectrum can not be universally described by a power-law, even of varying index. We have for example from Peebles (1993, p626) that 
\be
P(k)=\frac{(Q/A)A k}{(1+A k+B k^2)^2},\label{PS}
\ee
where
\bea
A\approx &\frac{8}{\Omega h^2}Mpc\approx 15.4Mpc,\nonumber\\
B\approx & \frac{4.7}{(\Omega h^2)^2}Mpc^2\approx 17.5 Mpc^2,\label{CDMspect}
\eea
and $Q$ is a scaling constant. If for convenience we choose to represent this as 
\be
P(k)=\frac{Q^*}{A}(Ak)^n,
\ee
so that 
\be
n(k)=1-\frac{2\ln{(1+Ak+Bk^2)}-\ln{(Q/Q^*)}}{\ln{Ak}},\label{nindex}
\ee
then we see that one can only have a smooth adiabatic variation in $n$ if one  sets 
\be
\ln{Q/Q^*}=2\ln{(2+B/A^2)}.
\ee
This ensures a smooth passage through $kA=1$. We observe that as $k\rightarrow\infty$ we have $n\rightarrow -3$ slowly, while it goes slowly to $n=1$ as $k\rightarrow 0$. Thus equation (\ref{an}) already ensures a flattened core and a power law density  $D_\rho=12/(\delta+1)$ at infinity. 

But equation (\ref{an}) also allows the density profile to be calculated in detail as a function of $k\approx 2\pi/r$, and consequently the variation with radius. Moreover  $\phi\approx \rho/(\sqrt{G\rho r^2})^3$ gives $D_\phi(r)=-D_\rho(r)/2+3$ as in the preceding section, so that this is also calculable. Moreover appealing to the general self-similarity relations found for example in H07, Henriksen (H06, 2006b) and in Henriken and Widrow (1999; $\delta$ there should be read as $1/a=2/D_\rho$) we have  for specific angular momentum, mass and radius respectively
\bea
j^2=Z& r^{(4-D_\rho)},\nonumber\\
M ={\cal M}(X)(\frac{r}{X})^{3-D_\rho},\\
X=\frac{r}{(\alpha t)^{2/D_\rho}}.\nonumber
\eea 
Here $X$, $Z$ (not related to the partition function of the previous section) and ${\cal M}(X)$ should be regarded as constants. Strictly speaking they are Lagrangian labels for different regions of the halo (just as are comparable factors in the density and pseudo-density), but in the spirit of adiabatic self-similarity used as a description of a self-similar cascade (see e.g HT84) they may also be regarded as constant between the scales. All of these variations may be found based simply on equation (\ref{an}).

We therefore define an adiabatically varying index for $j^2$ as 
\be
D_{j^2}(k)=-4+D_\rho(k).\label{Dj2}
\ee
We see moreover that near $r_s$ where $D_\rho\approx 2$, it is clear that $M_s\propto r_s$ and that $r_s\propto t$, that is to the age of the system. Similar relations are found in simulations (e.g. S07; Zhao et al. 2003; MacMillan, 2006;  MacMillan, Widrow and Henriksen, 2006).

\begin{figure}[ht!]
\begin{tabular}{cc} %This will make a two-column figure
\rotatebox{0}{\scalebox{.4} %change the angle and scale as you need
{\includegraphics{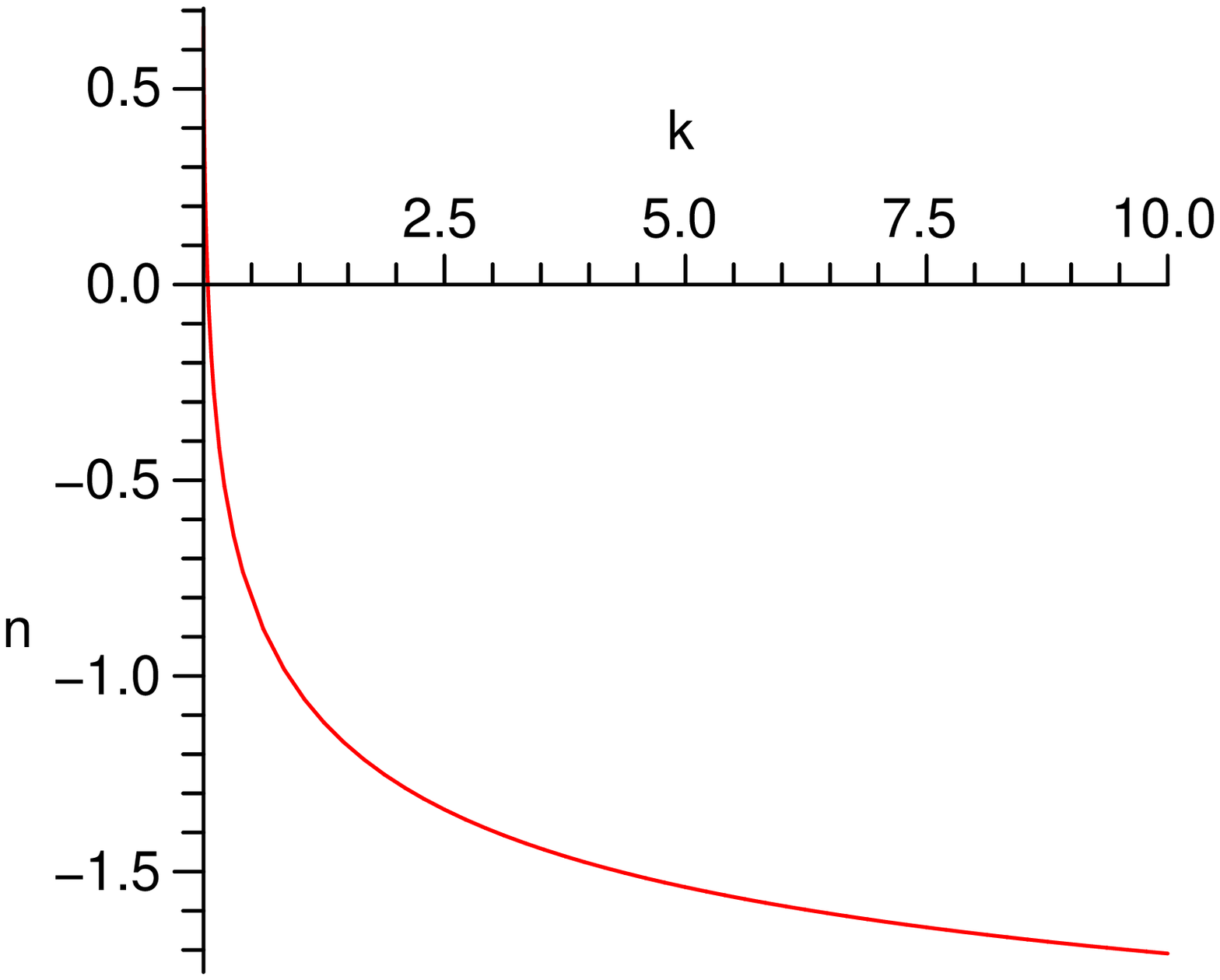}}}&
\rotatebox{0}{\scalebox{.4} %change the angle and scale as you need
{\includegraphics{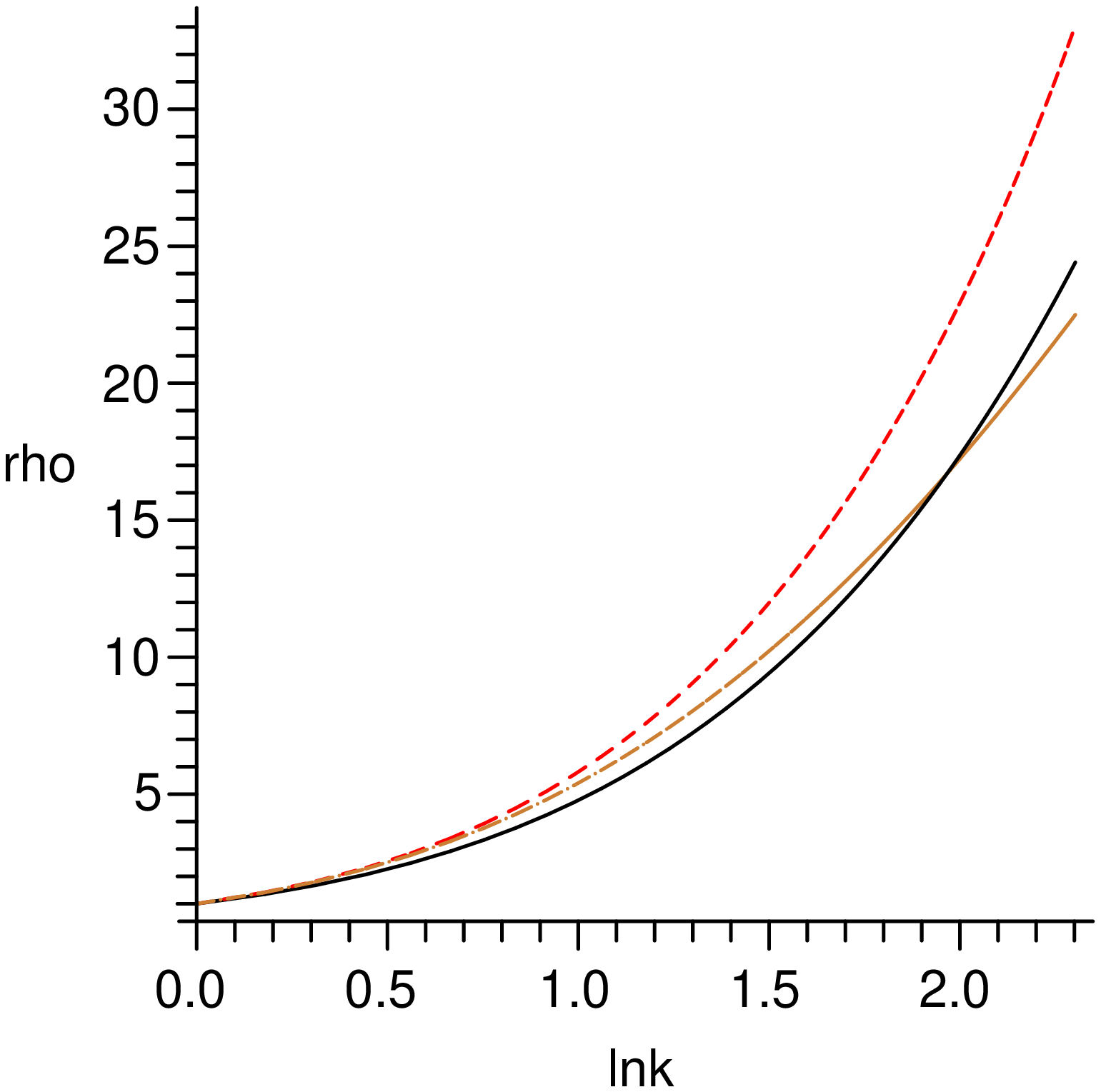}}}\\
\rotatebox{0}{\scalebox{0.4}
{\includegraphics{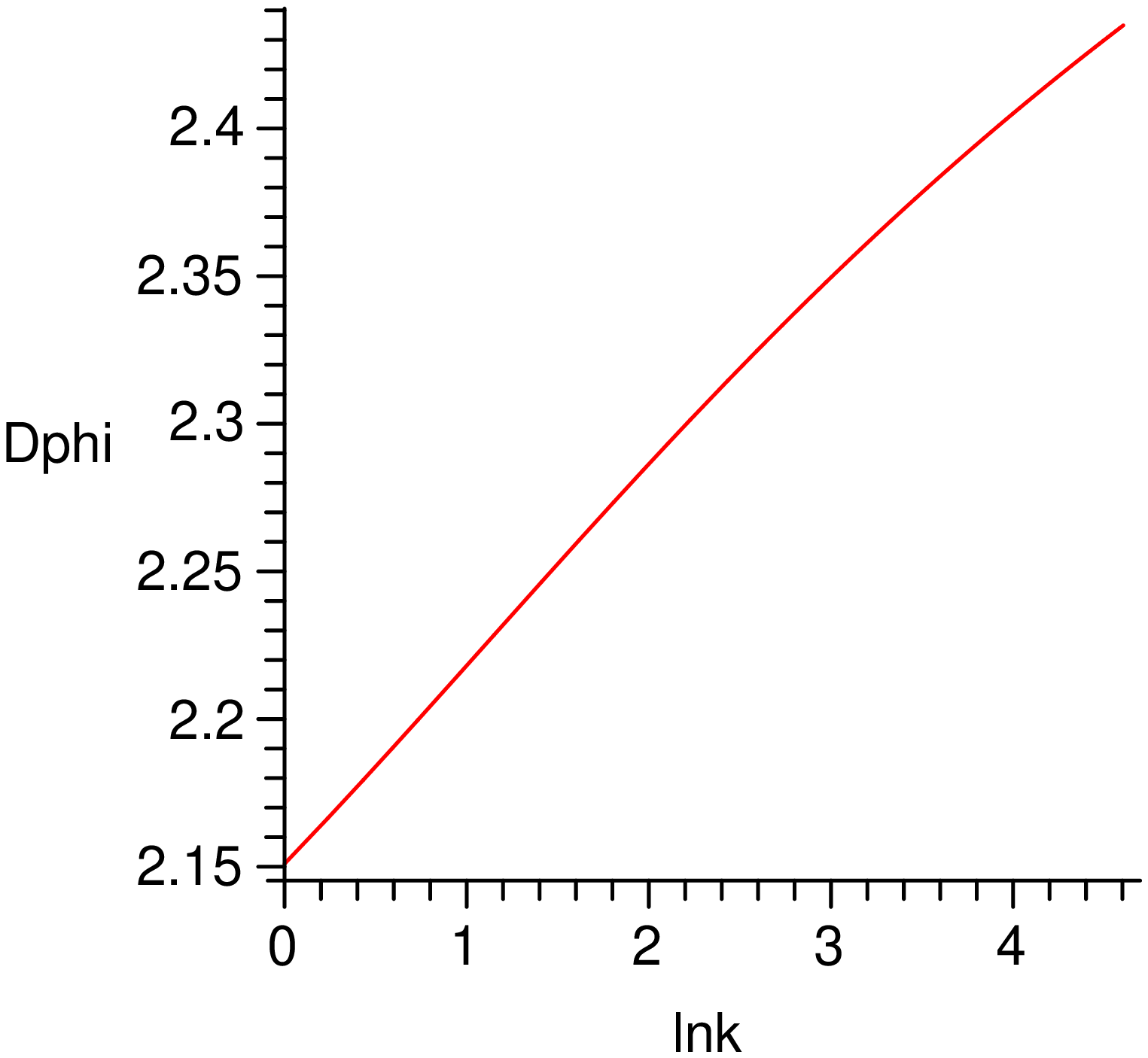}}}&
\rotatebox{0}{\scalebox{0.4}
{\includegraphics{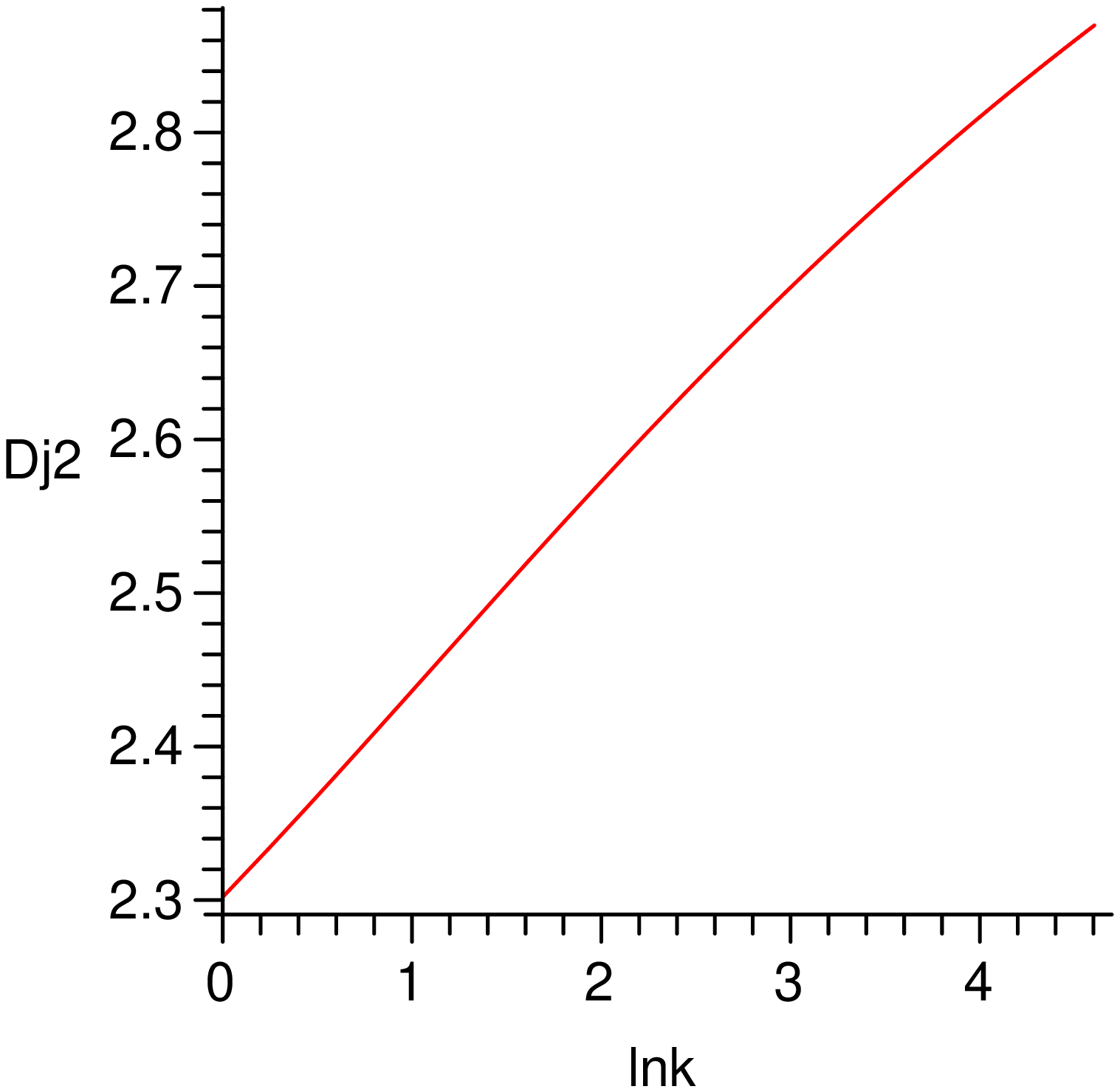}}}\\
\end{tabular}
\caption{
The top left panel shows $n(k)$ as found from equation (\ref{nindex}). The top right panel shows a comparison between $\rho(k)$ found from $n(k)$ and equation (\ref{an}) and the NFW profiles, all normalized at $k=1$. The middle curve is from Navarro et al. (2004), the upper curve is the original NFW profile and the bottom curve plots $\rho$ from the present calculation. The bottom left panel gives the index $D_\phi(k)$ while the bottom right shows $|D_{j^2}(k)|$. The calculations are presented for $\delta=4.5$.}
\label{fig:nature}
\end{figure}

This approach  yields an analytic construction of a dark matter halo, based ultimately on equation (\ref{an}). It develops that the calculation with $\delta=4.5$  gives a better fit to the NFW halos than either the case $\delta=5$ or $\delta=4$ so that we only show results for that case in figure (\ref{fig:nature}).
When $\delta=4$ the density increases too rapidly compared to the various NFW variations, while for $\delta=5$ it is too slow.

On the top right panel of figure (\ref{fig:nature}) we see how well this approach can be made to fit the empirical profile of Navarro et al. (2004) (the middle curve) while the upper dashed curve is the original NFW profile, which is too steep. The lower curve has been calculated using equations (\ref{an}) and (\ref{nindex}). This curve agrees in essence with S07 over this limited range. We see moreover that $D_\phi$ increases slowly towards the value $3$ once again while $D_{j^2}$ increases slowly towards $4$ as $k$ increases, both varying as $D_\rho\rightarrow 0$ with $r$. The mean rms tangential velocity thus becomes linear with radius in the core and rolls over to $\approx r^{0.15}$ near $r_s$. There is a remarkable similarity in the predicted evolution for $D_\phi$ and $Dj2$.

The interesting result is found in the comparison with figure (\ref{fig:evolve}). In that figure $w=-\ln{x}$ is basically indentical with $\ln{k}$ above. We see that there is qualitative agreement between the two approaches, but that the variations below $r_s$ occur much more rapidly in the maximum entropy approach. They occur in fact too rapidly to fit the numerical simulations over any substantial range (cf H07) if the relaxation begins close to $r_s$. This is less pronounced if the relaxation happens mainly at smaller scales. The divergence is best seen in the bottom right panel of figure (\ref{fig:evolve}). This shows the variation in $D_\rho$  for the present `nature' calculation, while the upper right panel shows it for the `nurture' calculation for a cascade beginning at $e^{-1}r_s$. Figure (\ref{fig:K2evolve}) shows the results when the cascade begins  at $e^{-2}r_s$, All variations are ultimately linear in the logarithm of scale, but the slopes are very different.

We must therefore conclude that this `nature' mixed with the adiabatic self-similar relaxation  describes the simulated halos very well, at least over a limited range in scales.  Maximizing the local entropy in the `nurture' approach can also be acceptable over a limited range in scale if the relaxation begins well below $r_s$. However it predicts a rapid ultimate divergence from the simulated behaviour.  There is no evidence for such `thermal relaxation' in the simulations at present. 

\section{Discussion and Conclusions}

The preceding sections allow us to conclude that the simulated dark matter halos are, up to the presently available resolution, consistent with the primarily `nature' (S07) adiabatically self-similar smooth growth from the CDM spectrum (e.g. \ref{fig:nature}). In our  semi-analytic treatment of the previous section, we imitate this growth using the notion of adiabatic  self-similarity (Henriksen, 2006b) to connect it to the CDM spectrum. In S07 however this was done using the assumption of smooth inside-out or `layer-cake' growth according to the Lacey and Cole (1993) prescription. This employs a Press-Schecter (1974) type argument for the CDM spectrum, together with the statistical treatment of Bond et al. (1991). We conclude that this may be done more simply using the notion of adiabatic self-similarity, which does  involve a measure of relaxation. Moreover, although it happens slowly, this approach predicts a central core ($n\rightarrow -3$) rather than a cusp.  

What we have termed variously `nurture',`thermodynamic' or `maximum entropy' relaxation based on a cascade of interacting structure (H07 and above) agrees on the qualitative trends with the `nature' view . However it evolves at small scales  towards a flat density and a steep pseudo-density more rapidly than is found either in the `nature' discussion or in the simulations if it is assumed to extend close to $r_s$ (\ref{fig:evolve}). This discrepancy is less pronounced when the relaxation extends only to smaller scales, however (e.g. figure(\ref{fig:K2evolve})). If we believe that cascade relaxation is playing a r\^ole in the halo evolution, then  equation (\ref{beta}) and the initial $\beta\approx -1$ tell us  that the relaxation  should indeed be most effective at small scales initially. In time, as $\beta$ becomes positive, the relaxation should move to large scales in agreement with equation(\ref{tJ}). 

The only relaxation that is visible in the simulations at present is in the density profile within two decades or so of $r_s$.  The phase space pseudo-density shows no relaxation over this same range. This behaviour can be fitted over  this limited range by either of the above relaxation mechanisms, but  both mechanisms predict  stronger small scale relaxation for which there is as yet no evidence in the simulations. 

One wonders whether the adiabatic self-similar relaxation that seems to fit  the current simulations well, is in fact due to a form of cascade relaxation. Recently a strong case has been made  (MacMillan, Widrow and Henriksen, 2006 (MWH06) and references therein) that the relaxation in this region near $r_s$ is due to the Radial Orbit Instability (ROI), so we should reconsider this in the present context.
 
The onset of this instability was found in the MWH06 paper to coincide with the development of the mean square specific angular momentum at $r$ into the Keplerian form $\propto GM(r)r$. Let us suppose that the angular momentum perpendicular to the radial direction takes the form $\ell_r^2\sigma^2(\ell_r)$, where we use the same notation as in equation (\ref{coupling}). Then the onset of the ROI is marked by the condition $\ell_r^2\sigma^2(\ell_r)\approx GM(r)r$ which on taking $\ell_r\approx r$ becomes essentially the cascade coupling condition of equation (\ref{coupling}). The explanation offered in the MWH06 paper is in terms of a resonant interaction between a `bar-like' density perturbation and `particle' (perhaps a sub-structure) orbits. In the present context we may see this as the largest asymmetric sub-structure at $r$. We  expect it to be composed of sub-structures and to be interacting with such objects in the environment. 

It s interesting to note incidentally that this interaction is a kind of coupling between the radial infall and the transverse orbital motion, which delivers free-fall energy to the top of the cascade. The Cascade is really  gravitationally driven turbulence. Ome might say that the mode $k\approx 1/r$ parallel to the infall is coupling to the transverse mode $k_r\approx 1/\ell_r$ perpendicular to the radial infall to produce this turbulence.  

Of course these are all  speculative, order of magnitude, arguments. Nevertheless we are inclined to suggest that the simulated halos, hence also the adiabatic self-similarity  relaxation that works well for the current simulations, is due to the ROI. This is in turn a mechanism for the conversion of radial infall into transverse orbital motion at $r$. This transverse mode represents the top of the cascade at each radius. We thus identify the ROI with a weak form of cascade relaxation at large scales. This weakness is in accord with equations (\ref{beta}) and (\ref{tJ}) that both predict ($\beta<0$ in the first case) stronger relaxation at small scales. The stronger cascade is expected to be at scales below the resolution of the current simulations.

We therefore conclude that cascade evolution may be at work both in reality and in the simulations. We conclude tentatively  that `Nature' type adiabatic self-similarity is the weak form of the cascade relaxation and that it is equivalent to the ROI. Our predictions that would confirm the presence of cascade evolution are that there is a halo density core rather than a cusp and that the pseudo-density power law should break at higher resolution in both scale and mass. In the Via Lactea run this region is just inside their reported convergence radius at about 1kpc. There may be weak evidence in figure 1 of DKM06 that a density flattening is not excluded.  Should this density and pseudo-density behaviour not appear down to say 100pc in a Milky Way type halo, then the strong cascade relaxation is not present in the simulations. It is then probably not present in reality, unless for some reason the necessary interactions at a distance between sub-structures is discriminated against in the simulations. DKM06 do observe that at lower resolution than that of the Via Lactea run, the small sub-halos are poorly resolved.

%We can however put the discrepancy in more cosmological terms. The discussion %can also  constrain the parameters of the structure cascade. Thus,  we note th%at the familiar expression for the correlation function $\xi(\ell)$ as an inte%gral over the power spectrum (e.g. Longair, 1998, p312),  should probably be w%ritten between the limits (a few times less than)$1/L$ and $1/\ell$ to give th%e correlation function of the $\ell$ substructures. 
%Together with $\ell/L=R$ constant, this permits (by the usual scaling of $k$) %extracting the adiabatically varying $n(k)$ from under the integral to deduce %as usual $\xi(\ell)/\xi(L)\propto (\ell/L)^{-(3+n)}$. However in the non-linea%r regime this is essentially $n(\ell,L)\propto (\ell/L)^{-D}$. Consequently, s%ince $D=2+\beta$ in our cascade model, we infer from $D=3+n$ that  initially $%n=\beta-1=(\nu-5)/2-\alpha=$ and after the evolution $n=-\alpha$. As discussed %previously we expect $\alpha\approx 0$. Thus we see from figures (\ref{fig:na%ture}, \ref{fig:evolve}) that the nurture evolution to the flat core at $n=0$ %occurs much more rapidly than the corresponding nature evolution towards a fla%t core at $n=-3$ for $K=1$. For $K=2$  so that the relaxation begins at smalle%r scales, this would be less pronounced over a limited range. 

%This argument does not change our previous conclusions, but it does however of%fer an explanation for the formation of the cascade of substructure inside 
%$r_s$ starting from a power spectrum with locally $n= -2-\alpha$, that is $\nu% =1$.   

\acknowledgements{
This work was supported in part by the Canadian Natural Science and Engineering Research Council. The author acknowledges helpful discussions with Eduardo Salvador-Sol\'e and Steen Hansen. This work grew out of the workshop on dark matter held at the Nils Bohr Institute in Copenhagen, August, 2007 }

%%%%%%%%%%%%%%%%%%%%%%%%%%%%%%%%%%%%%%%%%%%%%%%%%%%%%%%%%%%%%%%%%%%%%%%%%%
%%%%%%%%%%%%%%%%%%%%%%%%%%%%%%%%%%%%%%%%%%%%%%%%%%%%%%%%%%%%%%%%%%%%%%%%%%


\newpage

\begin{thebibliography}{}

\bibitem[An \& Wyn Evans(2005)]{An05} An, J. \& Evans N. W. 2005, A\&A, 444, 948
\bibitem[Austin \emp{et al} (2005)]{austin05} Austin, C.G., Williams, L.R., Barnes, E.I., Babul, A., \& Dalcanton, J.J. 2005, \apj, 634, 756

%\bibitem[Aguilar \& Merritt(1985)]{aguilar85} Aguilar, L. A. \&
%Merritt, D. 1985, \mnras, 217, 787

%\bibitem[Ascasibar \emp{et al.}(2004)]{asc04} Ascasibar, Y., Yepes,
%G., Gottlober, S., \& Muller, V.  2004, MNRAS 352, 1109

%\bibitem[Bardeen \emp{et al.}(1986)]{bbks} Bardeen, J.M., \emp{et
%al.} (BBKS) 1986, ApJ 304, 15

%\bibitem[Barnes \emp{et al.}(2005)]{barnes05} Barnes, E.I., Williams,
%L.L.R., Babul, A., \& Dalcanton, J.J 2005, ApJ, 634, 775

%\bibitem[Barnes \emp{et al.}(2006)]{barnes06} Barnes, E.I., Williams,
%L.L.R., Babul, A., \& Dalcanton, J.J. 2006, ApJ,643, 797

%\bibitem[Bertschinger(1985)]{bert85} Bertschinger, E.  1985, ApJS, 58,
%39

%\bibitem[Binney \& Tremaine(1987)]{bt87} Binney, J., \& Tremaine, S.
%1987, Galactic Dynamics (Princeton: Princeton Univ. Press)

%\bibitem[Bullock \emp{et al.}(2001)]{bull01} Bullock, J.S., \emp{et
%al.}  2001, ApJ 555, 240

%\bibitem[Carpintero \& Muzzio(1995)]{cm95} Carpintero, D.D., \&
%Muzzio, J.C.  1995, ApJ, 440, 5

%\bibitem[Carter \& Henriksen(1991)]{CH91} Carter, B., \& Henriksen, R. N. 1991%,J.Math.Phys., 32, 2580

\bibitem[Gonzalez-Casado \emp{et al.}(2007)]{GC07} Gonzalez-Casado, G., Salvador-Sol\'e, E., Manrique, A. \& Hansen, S.H. 2007, http://arxiv.org/abs/astro-ph/0702368
 
%\bibitem[Chen \& Jing(2002)]{chen02} Chen, D.N., \& Jing, Y.P.  2002,
%MNRAS, 336, 55

%\bibitem[Cho \emp{et al.}(2002)]{cho02} Cho, J., Lazarian, A. \& Vishniac, E.T%. 2002, ApJ, L566, 49

%\bibitem[Cincotta, N\'u\~nez, \& Muzzio(1996)]{cincotta96} Cincotta,
%P.M., N\'u\~nez, J.A., \& Muzzio, J.C.  1996, ApJ, 456, 274

%\bibitem[Col\'in \emp{et al.}(2000)]{colin00} Col\'in, P., Klypin,
%A.A., \& Kravtsov, A.V.  2000, ApJ, 539, 561

%\bibitem[de Vaucouleurs(1948)]{deV} de Vaucouleurs, G. 1948, 
%Ann. d'Astrophys., 11, 247

%\bibitem[Dehnen(2000)]{dehnen00} Dehnen, W.  2000, ApJL, 536, L39

%\bibitem[Dehnen(2001)]{dehnen01} Dehnen, W.  2001, MNRAS, 324, 273

%\bibitem[Dehnen(2005)]{dehnen05a} Dehnen, W.  2005, MNRAS, 360, 892

\bibitem[Dehnen \& McLaughlin(2005)]{dehnen05} Dehnen, W., \&
McLaughlin, D.E.  2005, MNRAS, 363, 1057

\bibitem[Diemand \emp{et al.}(2005)]{DMK06} Diemand, J., Kuhlen,M. \& Madau, P. 2006, \apj, 667, 859
 
%\bibitem[Dubinski \& Carlberg(1991)]{dub91} Dubinski, J., \& Carlberg,
%R.  1991, ApJ, 378, 496

%\bibitem[Evans \& Collett(1997)]{EC97} Evans, N. W., \& Collett, J.L. 1997, \a%pj, 480, L103

\bibitem[El-Zant \emp{et al}(2004)]{EZ04} El-Zant, A.A., Hoffman, Y., Primack, J., Combes, Francoise \& Shlosman, I. 2004, \apj,  607 , L75

\bibitem[Fillmore \& Goldreich(1984)]{fg84} Fillmore, J.A., \&
Goldreich, P.  1984, ApJ, 281, 1

%\bibitem[Frenk \emp{et al.}(1988)]{frenk88} Frenk, C.S., White,
%S.D.M., Efstathiou, G.P., \& Davis, M.  1985, Nature, 317, 595

%\bibitem[Frenk \emp{et al.}(1985)]{frenk85} Frenk, C.S., White,
%S.D.M., Davis, M., \& Efstathiou, G.P.  1988, ApJ, 327, 507

%\bibitem[Fridman \& Polyachenko(1984)]{fridman} Fridman, A. M. \&
%Polyachenko, V. L. 1984, {\it Physics of Gravitating Systems}
%(New York: Springer)

%\bibitem[Fukushige \& Makino(2001)]{fm01} Fukushige, T., \& Makino, J.
%2001, ApJ, 557, 533

%\bibitem[Ghigna \emp{et al.}(2000)]{ghigna00} Ghigna, S., Moore, B.,
%Governato, F., Lake, G., Quinn, T., \& Stadel, J.  2000, ApJ, 544, 616

%\bibitem[Goldreich \& Sridhar(1995)]{GS95} Goldreich, P. \& Sridhar, S. 1995, %\apj, 438,763

%\bibitem[Graham et al.(2005)]{graham05} Graham, A.W., Merritt, D., Moore, B., %Diemand,J. \& Terzi\'c, B., 2005, astro-ph/0509417

%\bibitem[Gunn(1977)]{gunn77} Gunn, J.E.  1977, ApJ, 218, 592

%\bibitem[Gunn \& Gott(1972)]{gg72} Gunn, J.E., \& Gott, J.R.  1972,
%ApJ, 176, 1

\bibitem[Hansen\emp{et al.}(2006)]{hansen06} Hansen, S., Moore, B., Zemp, M. \& Stadel J. 2005, Journal of Cosmology and Astroparticle Physics, 1, 14, also http://arxiv.org/abs/astro-ph/0505420

%\bibitem[H\'{e}non(1973)]{henon} Henon, M. 1973, A\&A, 24, 229

\bibitem[Henriksen\& Turner (1984)]{HT84} Henriken, R.N. \& Turner, B.E. 1984,
\apj, 287,200
\bibitem[Henriksen(1991)]{H91} Henriksen, R.N. 1991, \apj, 377,500

%\bibitem[Henriksen(1997)]{henrik97} Henriksen, R.N. 1997,in \lq Scale Invariance and Beyond'', Dubrulle, B., Graner, F.\& Sornette,  D.,(eds), Les Houches Workshop, Springer, Berlin,63

%\bibitem[Henriksen \& Widrow(1995)]{hw95} Henriksen, R.N., \& Widrow,
%L.M.  1995, \mnras, 276, 679

%\bibitem[Henriksen \& Widrow(1997)]{hw97} Henriksen, R.N., \& Widrow,
%L.M.  1997, \prl, 78, 3426

\bibitem[Henriksen \& Widrow(1999)]{hw99} Henriksen, R.N., \& Widrow,
L.M.  1999, MNRAS, 302, 321

%\bibitem[Henriksen \& Le Delliou(2002)]{hld02} Henriksen, R.N., \& Le Delliou,
%M. 2002, \mnras, 331,423

\bibitem[Henriksen(2006a)]{henrik06} Henriksen, R.N. 2006a, \mnras, 366, 697

\bibitem[Henriksen(2006b)]{H06} Henriksen, R.N. 2006b,\apj, 653, 894

\bibitem[Henriksen(2007)]{H07} Henriksen, R.N. 2007, \apj, 671, 1147
 
\bibitem[Hoffman \& Shaham(1985)]{hs85} Hoffman, Y., \& Shaham, J.
1985, ApJ, 297, 16

\bibitem[Hoffman \emp{et al.}(2007)]{hs07} Hoffman, Y., Romano-D\'iaz, E., Shlosman, I. \& Heller,C. 2007, \apj, 671,1108

%\bibitem[Huss, Jain, \& Steinmetz(1999)]{hjs} Huss, A., Jain, B., \&
%Steinmetz, M.  1999, ApJ, 517, 64

%\bibitem[Kazantzidis, Zentner, \& Kravtsov(2005)]{kazantzidis}
%Kazantzidis, S., Zentner, A. R., \& Kravtsov, A. V. 2006, \apj, 
%{\it in press}

%\bibitem[Kulessa \& Lynden-Bell(1992)]{kl92} Kulessa, A.S.\& Lynden-Bell, D. 1%992,\mnras,  255, 105

%\bibitem[Kunihiro(1995)]{K95} Kunihiro, T. 1995, Prog. Theor. Phys., 94, 503

\bibitem[Lacey\&Cole(1993)]{LC93} Lacey, C.\& Cole S. 1993, \mnras, 262, 627 

\bibitem[Le Delliou \& Henriksen(2003)]{ledelliou03} Le Delliou, M.,
\& Henriksen, R.N.  2003, A\&A, 408, 27

\bibitem[Longair(1998)]{Long98} Longair, M.S. 1998, {\it Galaxy formation}, Springer, Berlin-Heidelberg

%\bibitem[Lu \emp{et al.}(2006)]{lu} Lu, Y., Mo, H. J., Katz, N.,
%\& Weinberg, M. D. 2006, \mnras, 368, 1931

%\bibitem[Lynden-Bell(1967)]{ldb} Lynden-Bell, D. 1967, \mnras, 136, 101

\bibitem[MacMillan(2006)]{MacM06} MacMillan, J. 2006, PhD thesis, Queen's University at Kingston, Ontario, ON K7L 3N6, Canada

\bibitem[MacMillan, Widrow \& Henriksen(2006)]{mwh06} MacMillan, J.D., Widrow, L.M. \&
Henriksen, R.N., 2006, \apj, 653, 43

\bibitem[Madau \emp{et al.}(2008)]{MDK08} Madau, P., Diemand, J. \& Kuhlen, M. 2008, arXiv, 0802.2265M

\bibitem[Manrique \emp{et al.}(2003)]{M03} Manrique, A., Raig, A. Salvador-Sol\'e, E., Sanchis, T. \& Solanes, J.M. 2003,\apj,593,26

\bibitem[Merrall \& Henriksen (2003)]{MH03} Merrall, T.C. \& Henriksen, R.N. 2003, \apj, 595,43

%\bibitem[McDonald(2006)]{McD06} McDonald, P. 2006, Phys. Rev. D, 74,103512

%\bibitem[McDonald(2007)]{McD07} McDonald, P. 2007, Phys. Rev. D. 75, 043514
 
%\bibitem[May \& van Albada(1984)]{may84} May, A. \& van Albada,
%T. S. 1984, \mnras, 209, 15

%\bibitem[Merritt\& Aguilar(1985)]{me85} Merritt, D. \& Aguilar, L. A., 1985, \%mnras,217,787

%\bibitem[McDonald (2006)]{McD06} McDonald, P. 2006, astro-ph 0606028

%\bibitem[Moore \emp{et al.}(1998)]{moore98} Moore, B., Governato, F.,
%Quinn, T., Stadel, J., \& Lake, G.  1998, ApJ, 499, L5

\bibitem[Navarro, Frenk, \& White(1997)]{nfw} (NFW) Navarro, J.F.,
Frenk, C.S., \& White, S.D.M.  1997, ApJ, 490, 493

\bibitem[Navarro \emp{et al.}(2004)]{navarro04} Navarro, J.F.,
Hayashi, E., Power, C., Jenkins, A.R., Frenk, C.S., White, S.D.M.,
Springel, V., Stadel, J., \& Quinn, T.R.  2004, MNRAS, 349, 1039

%\bibitem[Nusser(2001)]{nusser01} Nusser, A.  2001, MNRAS, 325, 1397

%\bibitem[Palmer \& Papaloizou(1987)]{pp87} Palmer, P.L., \&
%Papaloizou, J.  1987, MNRAS, 224, 1043

%\bibitem[Peacock(1999)]{peacock99} Peacock, J. 1999, {\it Cosmological Physics%},
%Cambridge University Press, Cambridge, U.K.

%\bibitem[Peebles(1984)]{peebles84} Peebles, P. J. E. 1984, 
%\apj, 277, 470

\bibitem[Peebles(1993)]{peebles93} Peebles, P.J. E. 1993, {\it Principles of 
Physical Cosmology},  Princeton University Press, Princeton, NJ

%\bibitem[Polyachenko(1981)]{pol81} Polyachenko, V.L., 1981, Soviet Astr.Let., %7, 79

%\bibitem[Power \emp{et al.}(2003)]{power03} Power, C. \emp{et al.}
%2003, MNRAS, 338, 14

\bibitem[Press\&Schechter1974]{PS74} Press, W.H.\& Schechter, P., 1974, \apj, 187, 425

%\bibitem[Quinn, Salmon, \& Zurek(1986)]{quinn86} Quinn, P.J., Salmon,
%J.K., \& Zurek, W.H.  1986, Nature, 322, 329

%\bibitem[Ryden \& Gunn(1987)]{rg87} Ryden, B.S., and Gunn, J.E. 1987,
%ApJ 318, 15

%\bibitem[Ryden(1993)]{ryden93} Ryden, B.S. 1993,
%\apj 418, 4

%\bibitem[Shen \& Sellwood(2004)]{shen04} Shen, J., and Sellwood, J.A.
%2004, ApJ 604, 614

%\bibitem[Sikivie, Tkachev, \& Wang(1997)]{sikivie97} Sikivie, P.,
%Tkachev, I. I., and Wang, Y. 1997, \prd, 56, 1863

\bibitem[Salvador-Sol\'e(2007)]{S07} Salvador-Sol\'e, E., Manrique, A., Gonzalez-Casado, G., Hansen, S.H. 2007, ApJ, 666, 181

\bibitem[Taylor \& Navarro(2001)]{taylor01} Taylor, J.E., \& Navarro,
J.F.  2001, ApJ, 563, 483

\bibitem[Wyn Evans \& An(2005)]{NW-E05} Wyn Evans, N. \& An, J. 2005, \mnras, 360, 492

%\bibitem[Tremaine et al(1994)]{Tretal} Tremaine, S., Richstone, D.O., Yong-Ik %Byun, Dressler, A., Faber,S.M., Grillmair, C., Kormendy, J.\& Lauer, T.R. 1994%,\aj, 107,634

%\bibitem[van Albada(1982)]{van82} van Albada, T. S. 1982, \mnras, 201, 939

%\bibitem[Zaroubi, Naim, \& Hoffman(1996)]{zaroubi96} Zaroubi, S.,
%Naim, A., \& Hoffman, Y.  1996, ApJ, 457, 50

%\bibitem[Zeldovich(1970)]{zeldovich70} Zeldovich, Ya. B.  1970, A\&A, 5, 84

\bibitem[Zhao \emp{et al.}(2003)]{Zhao03} Zhao,D.H., Mo, H.J., Jing, Y.P. \& B\"orner, G. 2003,  \mnras, 339,12

\end{thebibliography}
\end{document}